\documentclass[12pt,a4]{article}
\usepackage{amsmath}
\usepackage{bm}
\usepackage{amsfonts}
\usepackage{amssymb}
\usepackage{graphics}
\usepackage[normal]{caption2}
\usepackage{subfigure}
\usepackage{rotating}
\usepackage{citesort}
\def\be{\begin{equation}}
\def\ee{\end{equation}}
\def\ba{\begin{array}}
\def\ea{\end{array}}
\def\bea{\begin{eqnarray}}
\def\eea{\end{eqnarray}}

\begin{document}
\baselineskip 20pt \setlength\tabcolsep{2.5mm}
\renewcommand\arraystretch{1.5}
\setlength{\abovecaptionskip}{0.1cm}
\setlength{\belowcaptionskip}{0.5cm}
\begin{center} {\large\bf THEORETICAL STUDIES OF RARE-EARTH NUCLEI LEADING TO
~$_{50}$Sn-DAUGHTER PRODUCTS
AND THE ASSOCIATED SHELL EFFECTS}\\
\vspace*{0.4cm}
{\bf Sushil Kumar}
\footnote{Email:~sushilk17@gmail.com}\\
$^a${\it  Department of Applied Science, Chitkara University, Solan
-174103,(H.P.) India.\\} 

\end{center}
Cluster decays of rare-earth nuclei are studied with a view to look for neutron magic shells for the $_{50}$Sn nucleus as
the daughter product always. The $^{100}$Sn and $^{132}$Sn radioactivities are studied to find the most probable cluster
decays and the possibility, if any, of new neutron shells. For a wide range of parent nuclei considered here (from
Ba to Pt) $^{12}$C from $^{112}$Ba and $^{78}$Ni from $^{210}$Pt parent are predicted to be the most probable clusters
(minimum decay half-life) referring to $^{100}$Sn and $^{132}$Sn daughters, respectively. Also, $^{22}$Mg decay of
$^{122}$Sm is indicated at the second best possibilty for $^{100}$Sn-daughter decay. In addition to these well known
magic shells (Z=50, N=50 and 82), a new magic shell at Z=50, N=66 ($^{116}$Sn daughter) is indicated for the $^{64}$Ni
decay from $^{180}$Pt parent.

\newpage
\baselineskip 20pt
\section{Introduction}
Since the discovery of $^{14}$C-decay from $^{223}$Ra by Rose and Jones \cite{rose84} in 1984, many other $^{14}$C-decays
from other radioactive nuclei ($^{221}$Fr, $^{221,222,224,226}$Ra, $^{223,225}$Ac and $^{226}$Th) and some 12 to 13
neutron-rich clusters, such as $^{20}$O, $^{23}$F, $^{22,24-26}$Ne, $^{28,30}$Mg and $^{32,34}$Si, have been observed
experimentally for the ground-state decays of translead $^{226}$Th to $^{242}$Cm parents
\cite{gupta94,bonetti99,bonetti07,gugliel08}, which all decay with the doubly closed shell daughter $^{208}$Pb (Z=82,
N=126) or its neighboring nuclei. Theoretically, such an exotic natural radioactivity of emitting particles (nuclei)
heavier than $\alpha$-particle was already predicted in 1980 by S\v andulescu, Poenaru and Greiner \cite{sandu80} on the
basis of the quantum mechanical fragmentation theory (QMFT) proposed by \cite{maruhn74,gupta75}.
Todate, $^{34}$Si is the heaviest cluster observed with the longest decay half-life ever measured
($log_{10}T_{1/2}(s)=29.04$) from $^{238}$U parent \cite{bonetti08}.
Recently, Poenaru et.al., extended the region of possible emitted clusters $A_c = 14 -34$
measured in the region of emitters with $Z$ = 87 - 96 to superheavy elements up to 124 \cite{poenaru11}.
In this systematic heavy particle radioactivity they consider not only the emitted cluters with atomic numbers
$2 < Z_{c} < 29$ but also heavier ones up to $Z_{c} = Z-82$, around $^{208}$Pb a doubly magic daughter.
For this purpose they used Analytical Superasymetric Fission (ASAF) model and estimated the half- life
for $^{128}$Sn emission from $^{256}$Fm (Q-value = 252.129 MeV) and for $^{130}$Te emission from $^{262}$Rf
(Q-value = 274.926 MeV): $log_{10}T^{Fm}(s)$ = 4.88 and $log_{10}T^{Rf}(s) $= 0.53, respectively.
They are in agreement with experimental values for spontaneous fission \cite{dchoff}: 4.02 and 0.32, respectively.

Keeping in mind the doubly magic nature of the $^{208}$Pb daughter, a second island of heavy-cluster radioactivity was
predicted on the basis of analytical superasymmetric fission model
(ASAFM)\cite{poenaru93} and preformed cluster model (PCM)\cite{kumar94}, in the decays
of some neutron-deficient rare-earth nuclei in to $^{100}$Sn (Z=N=50) daughter
or a neighboring nucleus. Furthermore, Kumar et al., \cite{kumar94}
predicted another doubly closed $^{132}$Sn (Z=50,N=82) daughter radioactivity, for decays of
some selective neutron-rich rare-earth nuclei. More ecently, an unexpected increase
in E2 strengths has been reported between the midshell isotope
$^{116}$Sn (Z=50, N=66) and its lighter neighbor, $^{114}$Sn \cite{Walker}, also a new shell closure at 
N=90 is predicted for the $^{140}$Sn isotope on the basis of shell model calculations \cite{sarkar10}. 
Experimentally, several unsuccessful attempts \cite{oganess94,gugliel95,gugliel97,mazzo02}
have been made to measure the $^{100}$Sn-daughter radioactivity from the $^{114}$Ba parent
nucleus produced in $^{58}$Ni+$^{58}$Ni reaction. Instead, a new phenomenon of intermediate
mass fragments (IMFs, with $3\leq Z \leq 9$), also referred to as 'clusters' or 'complex fragments',
emitted from the excited compound nucleus, was also observed \cite{campo88}.
It is worth mentioning that intermediate mass fragments  are mostly found in reactions at 
intermediate incident energies where  colliding nuclei breaks into many pieces\cite{vermani}.\\

In this paper,heavy cluster emissions of rare-earth parents (329 cases) with $_{50}$Sn always as the daughter
product is considered. The new experimental mass table \cite{audi03} and the theoretical masses \cite{moeller95}
are used to determine the released energy. Specifically, emission of various isotopes of C, O, Ne, Mg, Si, S, Ar,
Ca, Ti, Cr, Fe and Ni, are  considered respectively, from neutron-deficient
to neutron-rich Ba, Ce, Nd, Sm, Gd, Dy, Er, Yb, Hf, W, Os and Pt parents, with a view to look for $^{100}$Sn and $^{132}$Sn
radioactivities, as well as any other new Sn radioactivity with new shell closures in neutrons. Since the cluster
decays are more probable with daughters as magic nuclei, the decay half-lives are expected to drop
(be minimum) for the magic daughters. The same idea was utilized earlier
for the (spherical) sub-shell closed $_{40}$Zr daughter \cite{sushil03,sushil09}, including also a brief report of the
results on $_{50}$Sn daughter \cite{sushil03}. This calculation is based on the preformed cluster model (PCM) \cite{malik89,kumar97}, described briefly in Sect. II. The results of our calculation are presented in Sect. III and a summary of our results in Sect. IV.

\begin{figure*}
\includegraphics[width=15cm]{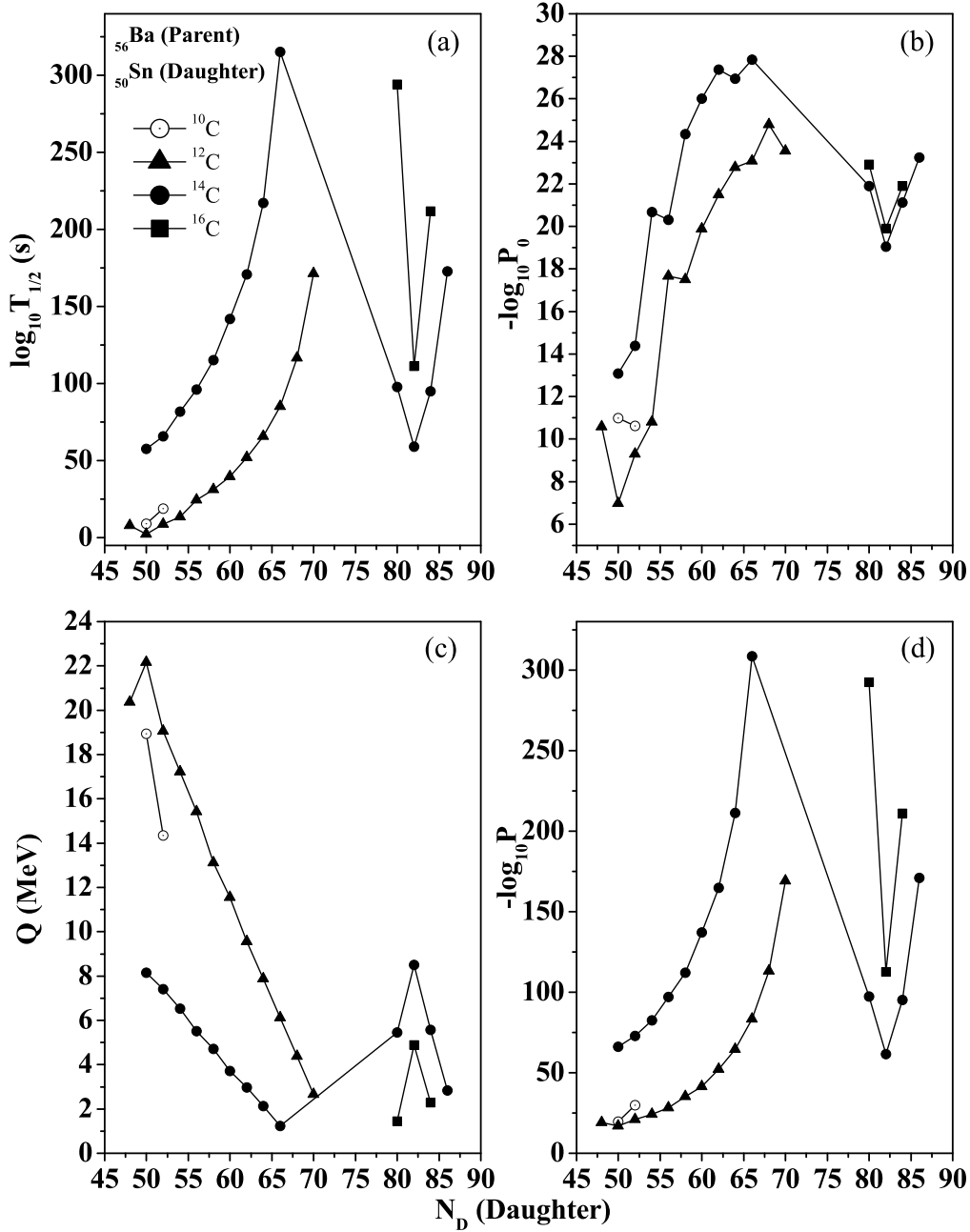}
\caption{The decay half-lives $T_{1/2}$(s) and other characteristic quantities like preformation factors $P_0$, Q-values
(in MeV), and penetrabilities $P$ of different Carbon clusters emitted with $_{50}$Sn daughters from various isotope of Ba
nuclei, calculated on the basis of the PCM, plotted as a function of daughter neutron number $N_{D}$.}
\end{figure*}

 \par
 \section{The model}
 The PCM model \cite{malik89} uses the dynamical collective coordinates of mass (and charge) asymmetry,
$\eta={{(A_1-A_2)}/{(A_1+A_2)}}$ and $\eta _Z={{(Z_1-Z_2)}/{(Z_1+Z_2)}}$,
first introduced in the QMFT \cite{maruhn74,gupta75}, which are in addition to the usual coordinates of
relative separation R and deformations $\beta_{2i}$ ($i=1,2$) of two fragments. Then, in the standard approximation of
decoupled R and $\eta $ motions, in PCM, the decay constant $\lambda$ or the decay half-life $T_{1/2}$ is defined as
\begin{equation}
\lambda ={{{ln 2}\over{T_{1/2}}}}=P_0P\nu _0,
\label{GrindEQ__1_}%
\end{equation}

 Here $P_0$ is the cluster (and daughter) preformation probability and $P$, the barrier penetrability, which refer to the
$\eta$ and R motions, respectively. $\nu _0$ is the barrier assault frequency.
The $P_0$ are the solutions of the stationary Schr\"odinger equation in $\eta$,
\begin{equation}
\Bigr \{ -{{\hbar^2}\over {2\sqrt B_{\eta \eta}}}{\partial \over {\partial
\eta}}{1\over {\sqrt B_{\eta \eta}}}{\partial\over {\partial \eta
}}+V_R(\eta )\Bigl \} \psi ^{({\omega})}(\eta ) = E^{({\omega})} \psi ^{({\omega})}(\eta ),
\label{GrindEQ__2_}%
\end{equation}
which on proper normalization gives
with $\omega$=0,1,2,3... . Eq. (\ref{GrindEQ__2_}) is solved at a fixed $R=R_a=C_t (=C_1+C_2)$ (the first turning point of the WKB integral, defined below), where $C_i$ are the S\"ussmann central radii $C_i=R_i-({1/R_i})$ (in fm), with the radii
$R_i=1.28A_i^{1/3}-0.76+0.8A_i^{-1/3}$. Many other radius formulas are available  \cite{ishwer} and widely used for the calculations of 
barrier heights, is also a subject of interest for the future study in PCM.\\
\begin{equation}
P_0={\sqrt {B_{\eta \eta}}}\mid \psi ^{({0})}%
(\eta (A_i))\mid ^2\left ({2/A}\right),\label{GrindEQ__3_}%
\end{equation}
The fragmentation potential $V_R(\eta )$ in (\ref{GrindEQ__2_}) is calculated simply as the sum of the Coulomb interaction, the
nuclear proximity potential \cite{blocki77} and the ground state binding energies of two nuclei,
\begin{equation}
V(R_a, \eta) =- \sum_{i=1}^{2} B(A_{i}, Z_{i})+ %
\frac{Z_{1} Z_{2} e^{2}}{R_a} + V_{P},\label{GrindEQ__4_}%
\end{equation}
The proximity potential between two nuclei is defined as 
\begin{equation}
V_{p}=4\pi\overline C\gamma b\Phi (\xi )%
\label{EQ_5_}%
\end{equation}
here $\gamma$ is the nuclear surface tension coefficient, $\overline C$ determines the distance between two points of the surfaces, evaluated at the point of closest approach  and $\Phi (\xi )$ is the universal function, since it depends only on
the distance between two
nuclei, and is given as
\(\begin{array}{clcr}
\Phi(\xi)=&
-0.5(\xi-2.54)^{2}- 0.0852(\xi-2.54)^{3},\\ 
& for\quad\xi\leq 1.2511\\
=&(-3.437 exp(-\xi/0.75).  
for\quad\xi\geq1.2511&%
\end{array}\)%

Here, $\xi$= s/b, i.e s in units of b, with the separation distance
s=$R-C_{1}-C_{2}$.
b is the diffuseness of the nuclear surface, given by
\begin{equation}
b=\left[\pi /2\sqrt{3} \ln 9\right]_{t_{10-90}}%
\label{EQ_7_}
\end{equation}
where $t_{10-90}$ is the thickness of the surface
in which the density profile changes from 90$\%$ to 10$\%$.
The $\gamma$ is
the specific nuclear surface tension, given by
\begin{equation}
\gamma =0.9517\left[1-1.7826\left(\frac{N-Z}{A} \right)^{2}%
\right] MeV fm^{-2}.%
\label{EQ_8_}%
\end{equation}
In recent years, many more microscopic potentials are available that takes care various aspects such as
overestimation of fusion barrier in original proximity potential, isospin effects. A comparison is also
available between all models \cite{puri10}. As noted above, even modified proximity potentials were also given.
We plan to study cluster decay with these new proximity potentials in near future. Here
 B's are taken from the 2003 experimental compilation of
Audi and Wapstra \cite{audi03} and, whenever not available in \cite{audi03}, from
the 1995 calculations of M\"oller {$\it et al.$} \cite{moeller95}. Thus, full shell effects are contained in our
calculations that come from the experimental and/or calculated binding energies.We also note that for exotic 
clusters/nuclei with neutron/proton rich matter, new binding energies are also available \cite{goyal}.
The momentum dependent potentials and symmtry energy potential which are found to have drastic effect at higher densities
will not affect decay studies, since these  happens at lower tale of the density \cite{sood,kumar}.
Here in Eq. (4), the Coulomb and proximity potentials are for spherical nuclei, and charges $Z_1$ and $Z_2$ in (4) are fixed by minimizing the potential in $\eta_Z$ coordinate. The mass parameters $B_{\eta \eta}(\eta )$, representing the kinetic
energy part in Eq. (\ref{GrindEQ__2_}), are the classical hydrodynamical masses of Kr\"oger and Scheid \cite{kroeger80}, used here for simplicity.

The penetrability $P$ is the WKB tunnelling integral, solved analytically \cite{malik89} for the second turning point
$R_b$ defined by $V(R_b)$=Q-value for the ground-state decay, and the assault frequency $\nu_0$ in (\ref{GrindEQ__1_}) is given
simply as
\begin{equation}
\nu_0=(2E_2/\mu )^{1/2}/R_0,\label{EQ__9_}%
\end{equation}
with $E_2=(A_1/A) Q$, the kinetic energy of cluster (the lighter fragment), for the Q-value shared between the two
products as inverse of their masses. $R_0$ is the radius of parent nucleus, and $\mu$, the reduced mass.
\begin{figure}
\includegraphics[width=15.0cm]{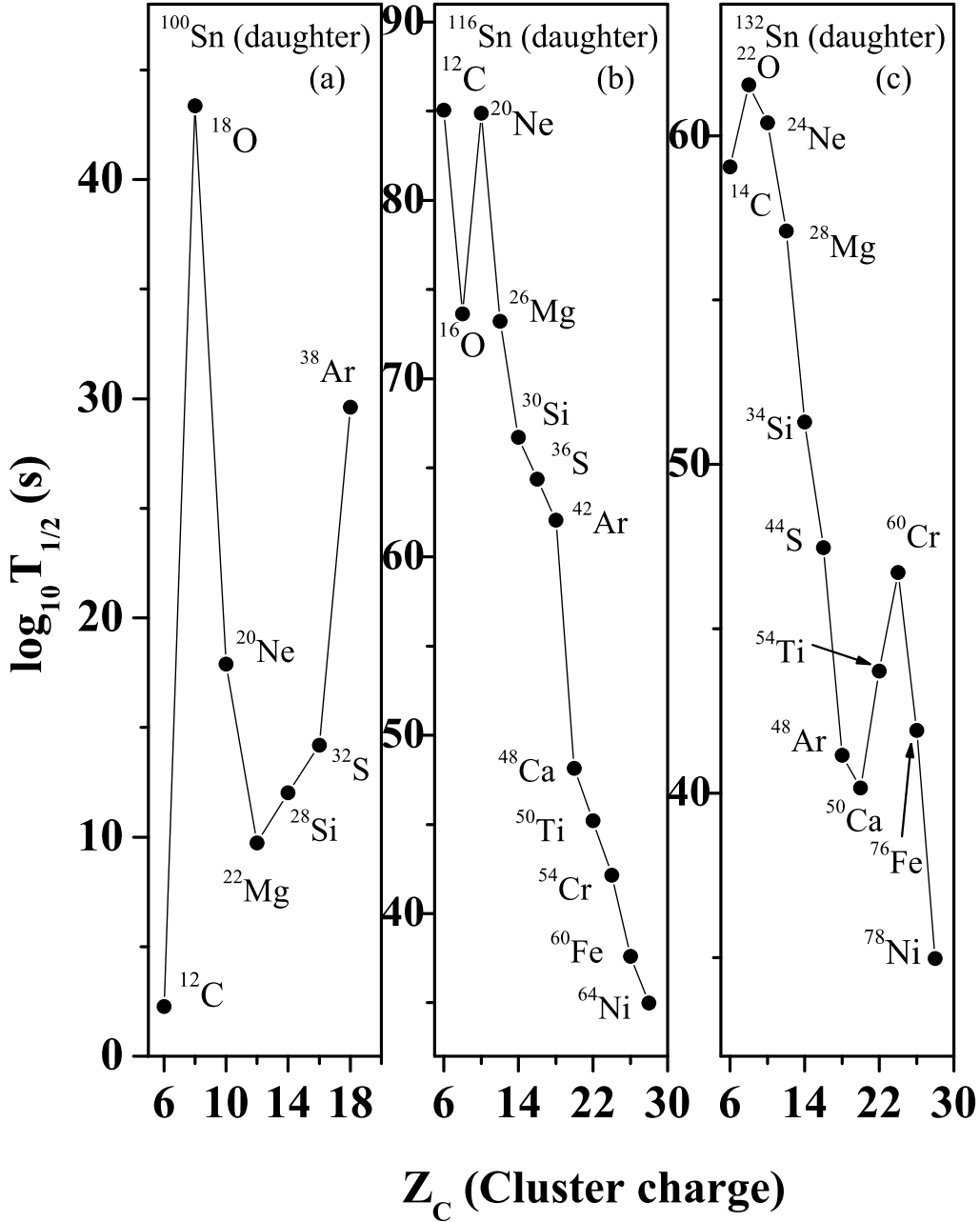}
\vskip-3mm\caption{$log_{10}T_{1/2}(s)$ for the most probable clusters emitted from various Ba to Pt parents with (a) $^{100}$Sn
(b) $^{116}$Sn and (c) $^{132}$Sn daughter, calculated on the basis of the PCM, plotted as a function of cluster proton
number $Z_{2}$. Note the different ordinate-scales used in these figures.}

\end{figure}

\par
 \section{Results and discussion}
As already stated in the introduction, the cluster decays of various isotopes of $_{56}$Ba to $_{78}$Pt parents are
calculated for the daughter nucleus to be always an isotope of $_{50}$Sn nucleus. For example, for the neutron-deficient
$^{110-132}$Ba and neutron-rich $^{144-150}$Ba parents considered here, different isotopes of Carbon cluster would give
rise to various isotopes of $_{50}$Sn daughter. This is illustrated in Fig. 1 for the decay half-life $T_{1/2}$ of
various C-decays, together with the Q-values, logarithms of penetrability P and preformation factor $P_0$, as a function
of $N_{D}$, the neutron number of $_{50}$Sn daughter. The impinging frequency $\nu_0$ is nearly constant
$\sim 10^{21}(s^{-1})$. All the four quantities Q, P, $P_0$, and $T_{1/2}$ show the shell effects at magic $N_{D}$=50 and
82; the Q, P and $P_{0}$ being large and $T_{1/2}$ small at these numbers. Thus, the most favorable decay is $^{12}$C from
$^{112}$Ba nucleus in the $48\leq  N_{D}\leq 70$ region, leaving behind $^{100}$Sn as the daughter product, and the $^{14}$C cluster from $^{146}$Ba in the $72\leq N_{D}\leq 86$ region with $^{132}$Sn as the daughter product. This result is same as in Refs. \cite{kumar94} where the most probable clusters for $^{100}$Sn daughters were obtained as the $A_{2}=4n$, N=Z, $^{12}$C, $^{16}$O, $^{20}$Ne, $^{24}$Mg and $^{28}$Si,
emitted from the respective Ba to Gd parents, and that these were the $^{14}$C, $^{20}$O, etc., for $^{132}$Sn daughter, 
emitted from $^{146}$Ba, $^{152}$Ce, etc.

In the present study, however, the other most probable clusters considered are (isotopes of O, Ne, Mg, Si, S, Ar, Ca,
Ti, Cr, Fe and Ni) from heavier neutron-deficient and neutron-rich rare-earth parents ($^{118-170}$Ce, $^{118-176}$Nd,
$^{122-184}$Sm, $^{132-190}$Gd, $^{132-194}$Dy, $^{138-200}$Er, $^{148-200}$Yb, $^{154-208}$Hf, $^{156-208}$W,
$^{160-210}$Os and $^{168-210}$Pt). Interestingly, $^{12}$C remains to be the most favorable cluster-decay from
$^{112}$Ba parent with $^{100}$Sn-daughter \cite{kumar94}, but for $^{132}$Sn-daughter the most favorable cluster is now
$^{78}$Ni from $^{210}$Pt, instead of $^{14}$C from $^{146}$Ba. This is illustrated in Fig. 2(a) and (c),
respectively, for $^{100}$Sn and $^{132}$Sn daughters, where the most probable clusters emitted from Ba to Pt parents are
plotted. The fact that the most probable cluster $^{78}$Ni, arising from Pt parents, occur at $N_{D}$=82 of the $_{50}$Sn
daughter is illustrated in Fig. 3 for $T_{1/2}$ alone. However, in Fig. 3, in addition to the strong minima at
$_{50}$Sn-daughter neutrons $N_{D}$=82, a new minimum is also shown to be persent at $N_{D}$=66 for the $_{50}$Sn-daughter,
emitting $^{64}$Ni cluster from $^{180}$Pt parent. This is further illustrated to be true in Fig. 2(b). Thus, a new
possibility of $^{116}$Sn-daughter radioactivity is indicated here. Apparently, other cases of interest in Fig. 2 are the
$^{22}$Mg decay of $^{122}$Sm and $^{50}$Ca decay of $^{182}$Yb, respectively, with $^{100}$Sn and $^{132}$Sn daughters.\\
\begin{figure}
\includegraphics[width=15.0cm]{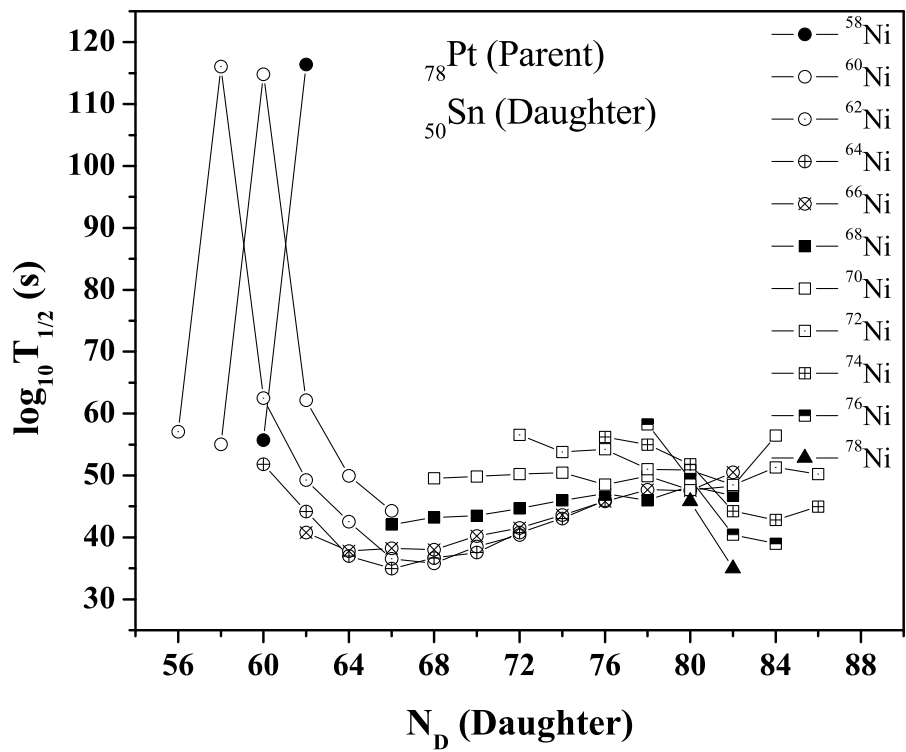}
\vskip-3mm\caption{Same as for Fig.1, but for $T_{1/2}$(s) alone, and for different Ni clusters emitted from
various Pt parents.}
\end{figure}

\begin{figure}
\includegraphics[width=15.0cm]{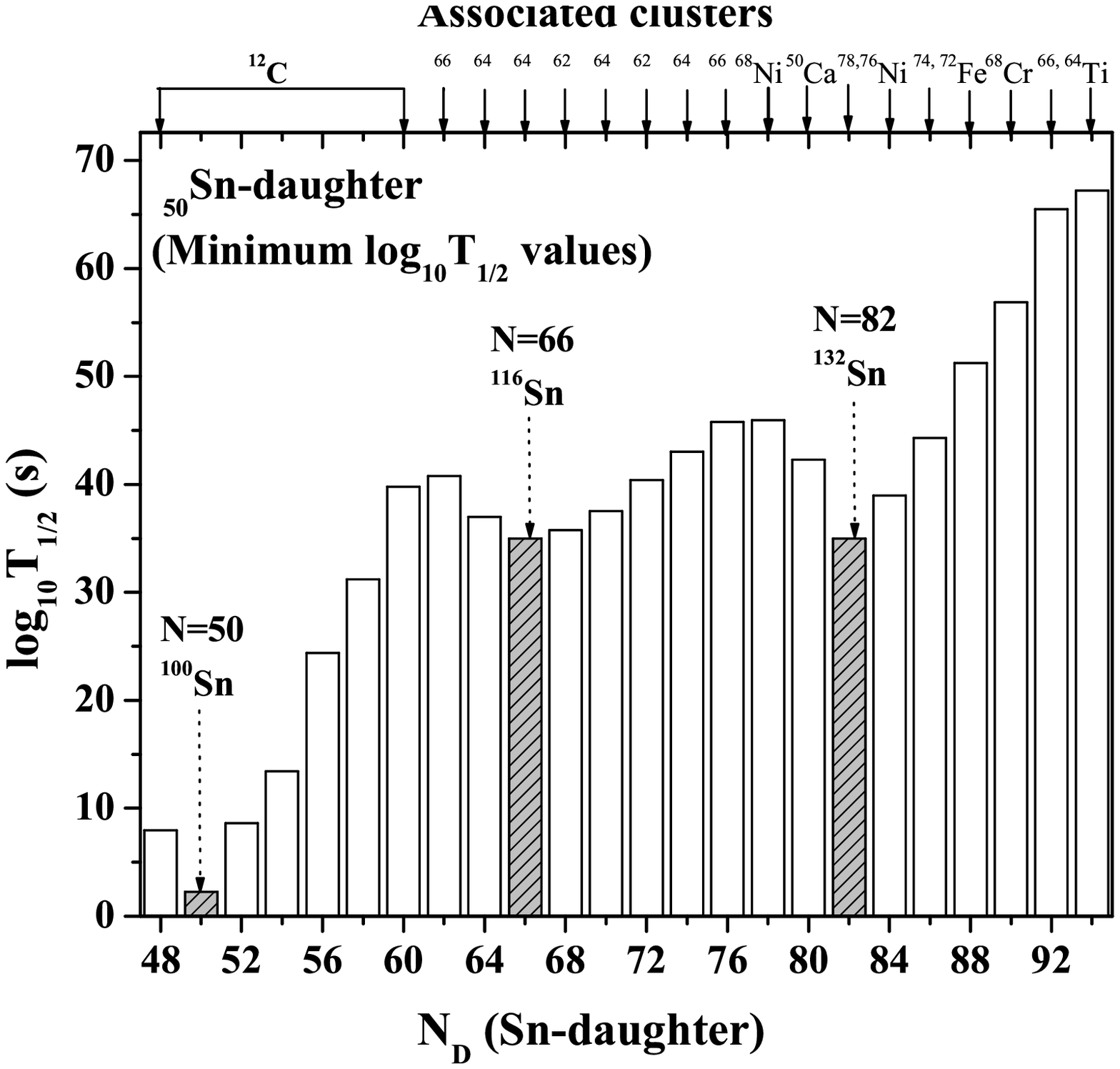}
\vskip-3mm\caption{Histogram of $log_{10}T_{1/2} (s)$ versus $_{50}$Sn-daughter neutron number $N_D$ for the most probable clusters
emitted from various Ba to Pt parents with $_{50}$Sn as a daughter nucleus always, calculated on the basis of the PCM.
The associated clusters are also shown on the top panel.}
\end{figure}

Finally, Fig. 4 gives a complete histogram of the decay half-lives $log_{10}T_{1/2}$(s) as a function of the neutron number
$N_{D}$ of the emitted $_{50}$Sn-daughters with the most probable clusters (minimum $T_{1/2}$ values) from some 329 parents
taken from Ba to Pt with mass numbers $A=110-210$. Limitation to $N_D \sim$94 since, for $N_D>$90, the
contribution from nuclei heavier than Pt would also become important. Note that in Fig. 4, $_{50}$Sn-daughter is kept fixed
and all possible clusters are considered from different parents (total 1617 combinations with $_{50}$Sn-daughter and the
probable cluster), and then the one with minimum half-life time is plotted. Apparently, the shortest half-life time
$log_{10}T_{1/2}(s)=2.27$ (with Q-value=22.16 MeV) is obtained for $^{12}$C decay of $^{112}$Ba. The role of magic
$N_{D}=82$ is also evident with a minimum in the histogram at $^{132}$Sn-daughter due to the emission of $^{78}$Ni cluster
from $^{210}$Pt parent. The predicted half-life $log_{10}T_{1/2}(s)=34.974$ (with a Q-value=119.292 MeV), which is beyond
the limit of present day experiments. Thus, as expected, the strongest shell effects occur at $N_{D}=50$ and 82. In
addition, another minimum due to $^{64}$Ni cluster emitted from $^{180}$Pt parents could also be of interest for a closed
shell (either spherical and/or deformed) at $N_{D}= 66$. This minimum is comparable to $N_{D}=82$ case, with a predicted
decay half-life also of nearly the same value ($log_{10}T_{1/2}(s)=34.975$, with a Q-value=124.192 MeV), which by all means
is again very large for experiments. Note from Fig. 2 that the decay half-life for $^{22}$Mg emitted from $^{122}$Sm
($log_{10}T_{1/2}(s)=9.735$) lies in between the values for $^{12}$C decay of $^{112}$Ba and $^{78}$Ni cluster from
$^{210}$Pt parent (or $^{64}$Ni cluster emitted from $^{180}$Pt parent), rather closer to the $^{12}$C decay of $^{112}$Ba.

\par
\section{Summary}
 The preformed cluster model (PCM) is used for the cluster decay calculations with $_{50}$Sn as a daughter nucleus always
from various parents of Ba to Pt region. Thus $^{100}$Sn and $^{132}$Sn-daughter radioactivities is
look for the most probable clusters (minimum decay half-life time) emitted from the rare-earth parents, and
the presence of any new neutron magicity. The most probable clusters, respectively, with $^{100}$Sn and $^{132}$Sn
daughters, are predicted to be $^{12}$C from $^{112}$Ba and $^{78}$Ni from $^{210}$Pt parent. Further possibilities with
$^{100}$Sn and $^{132}$Sn daughters are also noticeable in $^{22}$Mg and $^{50}$Ca clusters emitted from $^{122}$Sm and
$^{182}$Yb parents, respectively, as the second best new cases. In addition, a new shell is indicated at $N_{D}$=66  with
$^{116}$Sn-daughter due to $^{64}$Ni cluster emitted from $^{180}$Pt parents. However, at present these calculations seem
to be more of an academic interest since the predicted half-life times, for at least the $^{116}$Sn and $^{132}$Sn-daughter
radioactivities, are too large for experiments.

\par
The author is thankful to Prof. R. K. Gupta for many fruitfull discussion.

\end{document}